\documentclass[journal=jpcafh,manuscript=article]{achemso}

\usepackage{chemformula} 
\usepackage[T1]{fontenc} 
\usepackage{braket}
\usepackage{amsmath}
\usepackage{color}
\usepackage{ulem}

\author{Henning Kirchberg}
\affiliation{I. Institut f\"ur Theoretische Physik, Universit\"at Hamburg, Jungiusstr.\ 9, 20355 Hamburg, Germany}
\email{henning.kirchberg@physik.uni-hamburg.de}
\phone{+49-(0)40-42838-3012}
\author{Michael Thorwart}
\email{michael.thorwart@physik.uni-hamburg.de}
\affiliation{I. Institut f\"ur Theoretische Physik, Universit\"at Hamburg, Jungiusstr.\ 9, 20355 Hamburg, Germany}

\title[Time-Dependent Stokes-Shift]
 {Time-Resolved Probing of the Nonequilibrium Structural Solvation Dynamics by the Time-Dependent Stokes Shift}

\begin{document}

\begin{abstract}
The time-dependent fluorescence Stokes shifts monitors the relaxation of the polarization of a polar solvent in the surroundings of a photoexcited solute molecule, but also the structural variation of the solute following photoexcitation and the subsequent molecular charge redistribution. Here, we formulate a simple nonequilibrium quantum theory of solvation for an explicitly time-dependent continuous solvent. The time-dependent solvent induces nonequilibrium fluctuations on the solvent dynamics which are directly reflected in different time components  in the time-dependent Stokes shift. We illustrate the structural dynamics in the presence of an explicitly time-dependent solvent by the example of a dynamically shrinking solute which leads to a bi-modal Stokes shift. Interestingly, both contributions are mutually coupled. Furthermore, we can explain a prominent long-tail decay of the Stokes shift associated to slow structural dynamical variations.
\end{abstract}

\maketitle 

\section{Introduction}
\label{StokesShift}
A photoionized solvated molecule with its suddenly rearranged electronic charge distribution is no longer in equilibrium with its surrounding solvent. The subsequent electronic relaxation can be monitored spectroscopically by measuring the time-dependent Stokes shift. During the relaxation, the molecular structure is dynamically reorganized and multiple time scales may be involved in this structural dynamics of the solute-solvent system \cite{lag2017}. One time scale is set by the correlation time of the fluctuating electric field which is induced by the surrounding polar solvent molecules. Commonly, these fluctuations are described by thermal equilibrium variations about the local mean, which modify, e.g., the free energies of reactants, products and transition states and, thus, affect the energy of activation or even the course of the chemical process \cite{NitzanBook}. A further time scale may be associated to the intrinsic structural dynamics of the solute itself which may influence its electronic relaxation properties, in concert with the dynamical rearrangement of the solvent which can occur far from thermal equilibrium. Such a dynamical solvent induces then  nonequilibrium fluctuations of the electric field acting on the solute's charge configuration and their influence on the relaxation dynamics is more difficult to be included.

A large body of literature exits for the relaxational dynamics in solution due to equilibrium fluctuations of the solvent molecules. For instance,  Fleming and co-workers \cite{jim1994,ros1991} have uncovered that the solvation dynamics at very short times is associated to inertial collective relaxation of solvent molecules mostly driven by the large force constant of solvent free energy surfaces \cite{ros1991,lag2017}. The latter are also employed in the famous Marcus theory to describe solvent-assisted charge transfer reactions \cite{mar1964,mar1965}. Other ultrafast processes in water are the intramolecular O-H stretching (10 fs) and bending vibrations (20 fs), which occur on the time scale of a few tens of femtoseconds\cite{lag2017}.

 To probe molecular motion at the interfaces between the solute and the solvent on different time scales, time-dependent fluorescence spectroscopy is a common technique. By determining the Stokes shift, which is the shift of the maximum of the fluorescence emission spectrum a time span $t$ after initial excitation, solvent relaxation in the close proximity of the solute can be recorded for varying times  \cite{hei2019,nil2005,bag1984}. The time-dependent Stokes shift is commonly discussed in the context  when an electronic excitation occurs in the framework of the Franck-Condon principle \cite{fra1926,con1926}. According to this, the configuration of the polar solvent immediately after photoionization of the solute is still that of the solvent with the unexcited solute. For the total solute-solvent system,  this is an energetically unfavorable configuration, and in the course of time, the solvent adopts dynamically (different parts of it on different timescales) to the altered charge distribution of the solute. It reaches energetically more favorable configurations, so that the energy of the fluorescence emission decreases over time.
 
The time evolution of the Stokes shift in polar media at equilibrium is described by the well-established Bagchi-Oxtoby-Fleming theory in terms of a continuum Onsager model of the solvent \cite{bag1984}. It may be understood as a time-dependent description of the Ooshika-Lippert-Mataga equation of the average shift in frequency of the absorption and fluorescence transition in solution due to polar interactions \cite{bag1984,oos1954,mat1956,lip1957}. Using the theory of linear response, it is possible to link the fluctuations of the solute-solvent interaction to the dipole equilibrium correlation function of the solute alone, which depends on solvent time constants and dielectric properties only \cite{bag1984}. For the simplest case of a Debye-type dielectric medium \cite{deb1913}, the model predicts an exponential relaxation of the solvation energy with the time constant being given by the solvent's longitudinal relaxation time $\tau_L=[(2\epsilon_\infty+1)/(2\epsilon_s+1)]\tau_D$. In general, $\tau_L$ is strongly reduced in comparison to the Debye relaxation time $\tau_D$ for a solvent with a large static dielectric constant $\epsilon_s$ like water due to collective polarization dynamics. This becomes observable by the time-dependent Stokes shift \cite{bag1984}. 

The approach in terms of a continuum model used for time-resolved fluorescence spectroscopy is a powerful description until today because the fluorescence response is mostly insensitive to the motion of individual water molecules but only to their collective motion \cite{hei2019,nil2005}. Spectroscopic measurements confirm relaxation dynamics in accordance with $\tau_L$ which is much faster than the process of reorientation of individual solvent molecules.  This reflects the fact that the solvent response involves the concerted motion of many molecules. The prediction of $\tau_L$ forms a cornerstone for the comparison of experimental data of the time-dependent Stokes shift to theoretical models \cite{mar1988}. 

However, experimental data sometimes deviate from the predictions of a homogeneous continuum model. This may be understood from the microscopic picture. The solvent in close proximity to the solute contains an insufficient number of molecules to attain the full cooperativeness described by the $\tau_L$ response, but the solvent further away may look like a continuum fluid where the $\tau_L$ response pertains. This fact is known as the "Onsager inverse snowball picture", which associates the shorter time scales with solvent layers further away from the solute \cite{ons1977}. Therefore, relaxation may occur on several different time scales, ranging from single particle relaxation processes up to the collective polarization dynamics characterized by $\tau_L$. Even  ultrafast relaxation contributions characterized by a non-exponential decay may arise \cite{mar1988,bag1992}. However, Bagchi and Chandra have shown that when translational motions of the solvent become important, which are neglected in Onsager's approach,  relaxation is generated by different solvent shells such that the fastest dynamics is associated with the closest shells in the vicinity of the solute \cite{bag1988,bag1989,bag2012}. 

An alternative to a continuum description are molecular dynamics (MD) simulations or combined quantum mechanics/ molecular mechanics (QM/MM) simulations. They help to mimic at least the most prominent solvent reorientational and relaxational time scales to explain experimental time-dependent fluorescence studies. Neira and Nitzan, for example, have used MD simulations to confirm a slow component of the dynamical solvent  associated with salt ions in electrolyte solutions. Their signature becomes visible by a slower time component in the Stokes shift in comparison to the conventional bulk water relaxation \cite{ner1993}. 

Recent experimental time-dependent fluorescence studies of aqueous proteins provide evidence of different contributions to relaxation associated with distinct motions of such complex systems like proteins \cite{cer2019,hei2019}. Heid and Braun, for example, decompose the fluorescence Stokes shift into a water and a protein component by performing MD simulations at nine different sites of the protein in water \cite{hei2019}. The water component dominates the static Stokes shift at short times, but its magnitude decays rapidly. In turn,  the self-motion of the protein becomes visible after a few picoseconds. The resulting Stokes shift therefore leads to a bi- or even multimodal decay  \cite{hei2019,nil2005}. Despite the computational accessibility to describe different dynamical contributions of the solvation process, MD simulations are usually limited to a numerically tractable number of solvent molecules. Furthermore, simulations often use linear solvent response for a varity of pertubations such as changes of the solute's electronic state. By this, the determination of computationally expensive forces using excited-state interaction potentials can be avoided, since the impact of the perturbation can be calculated from the fluctuations of the equilibrium solute-solvent interaction. However, nonequilibrium simulations clearly identify "mechanical" relaxation due to changes in the spatial extension of the solute which are key to the breakdown of linear response. \cite{ahe2000}

In this work, we combine the self-motion of the solute with a continuum description of a time-dependent solvent and determine the resulting time-dependent Stokes shift. We formulate an effective nonequilibrium theory in which an explicit time-dependent motion of the continuum solvent beyond thermal fluctuations enters. By this, we rejuvenate and generalize the accostable Bagchi-Oxtoby-Fleming continuum theory by including an explicitly time-dependent Onsager model introduced in Ref.\ \cite{kir2018}. 
To be specific, we focus on the case of a dynamically shrinking radius of an Onsager sphere and determine fully analytically the time-dependent Stokes shift. This model combines nonpolar and dielectric relaxation dynamics and is an archetypal example for the solvation dynamics beyond Onsager's regression hypothesis which normally associates dynamical spectral signatures of the solute to spontaneous equilibrium fluctuations of the solvent. \cite{cha1987}
We find a bimodal decay in the time-dependent Stokes shift in which the shrinking rate of the solute and the correlation time of the solvent both appear.
Throughout the work, we concentrate on water as a solvent, but the theory applies to any other polar solvent.

\section{Model}
\label{StokesModel}
We consider the time-dependent Stokes shift characterized by the function \cite{NitzanBook} 
\begin{align}
\label{eq3.3.1}
S(t)=\frac{\Delta E_{solv}(t)-\Delta E_{solv}(\infty)}{\Delta E_{solv}(0)-\Delta E_{solv}(\infty)} 
\end{align} 
for a photoexcited molecular complex with explicit molecular motions at the interface to a dipolar solvent. Here,  
 $\Delta E_{solv}(t)$ is the resulting solvation energy difference between the excited and the ground state molecule at a given time $t$. It results from the electrostatic interaction between the charge distribution of the solute and the surrounding polar solvent. If there is only little internal vibrational excitation of the solvent during the initial photoexcitation, the time-dependent Stokes shift mainly results from the time-dependent solvation energy \cite{hsu1997}. 

To illustrate the mechanism of the structural dynamics of a coupled solute-solvent system, we assume that the self-motion of the solute is given by the model of a dynamically shrinking Onsager cavity \cite{ons1936,kir2018} in the center of which the central molecular dipole moment suddenly changes from an initial ground state dipole moment $\boldsymbol{\mu}_g$ to $\boldsymbol{\mu}_e$ upon photoexcitation which initiates the geometrical `shrinking'. The shrinking cavity radius may mimic generically motional changes at the solute-solvent interface such as the observed self-motion of dissolved proteins \cite{nil2005,hei2019}. Another example is the photodetachment of an electron from a sodium anion leaving behind a smaller neutral atom that drives solvent molecules into that locations which were occupied before by the volume of the solute. \cite{bed2003}

The electrostatic interaction energy of the dipole moment with the homogeneous and isotropic electric field provided by the polar solvent is given by  $V_I=-\boldsymbol{\mu}(t)\cdot \textbf{R}(t)\equiv E_{solv}(t)$. We disregard solute-solvent coupling of higher multipole contributions because the change upon excitation in the shape and size of the solute is usually small. The reaction field for a not to rapidly varying Onsager radius $a(t)$ according to Ref.\ \cite{kir2018} can be calculated to be 
\begin{align}
\label{eq1}
\textbf{R}(t)&=\frac{1}{a(t)^3}\int_{-\infty}^t dt' \frac{2(\epsilon_s-1)}{3\tau_D}\exp[-\omega_D (t-t')] \boldsymbol{\mu}(t')\\ \notag &=\frac{1}{a(t)^3}\int_{-\infty}^t dt' \chi(t-t') \boldsymbol{\mu}(t'),
\end{align}  
where $\omega_D=(2\epsilon_s+1)/(3\tau_D)=\tau_L^{-1}$. The static dielectric constant of water at 20$^{\circ}$C is $\epsilon_s=78.3$ and the Debye relaxation time $\tau_D=8.2$ ps \cite{mck2005}. The reaction field portrays the time-dependent back action of the polarization of the surrounding solvent due to the dipole on itself. We neglect in our model a possible change of the solute's polarizability upon excitation which may be included by additionally filling the Onsager cavity with a dielectric. A thorough discussion on this can be found in the work of Bagchi, Oxtoby and Fleming \cite{bag1984}. Here, we focus on structural changes of the solute at the solute-solvent interface which enter via a time-dependent Onsager radius $a(t)$. For an exponentially shrinking Onsager radius by a not too large amount from an initial value $a_0+a_1$ to a final value $a_0$,  we use the expansion $1/a^3(t)=1/(a_0+a_1\exp[-\alpha t]\Theta(t))^{3}\approx 1/a_0^{3}-3a_1\exp[-\alpha t]\Theta(t)/a_0^4$ up to first order in $a_1/a_0$. Here, $\alpha$ is the phenomenological shrinking rate. By this, we can split the reaction field of Eq.\ (\ref{eq1}) into two terms according to 
\begin{align}
\label{eq3.3.2}
\textbf{R}(t)=&\frac{1}{a_0^3}\int_{-\infty}^t d t' \bigg[\chi(t-t')-\frac{3a_1}{a_0}\exp[-\alpha t]\Theta(t)\chi(t-t') \bigg] \boldsymbol{\mu}(t'),
\end{align}
where the shrinking begins upon photoexcitation at time $t=0$. 
We assume the optical excitation of the solute to occur instantaneously such that the dipole moment changes from $\boldsymbol{\mu}_g$ to $\boldsymbol{\mu}_e$ at $t=0$ which coincides with the beginning of the radial shrinking. Moreover, we assume that the dipole moment does not change its direction but its magnitude according to
\begin{align}
\label{eq3.3.3}
\boldsymbol{\mu}(t)=\mu_g \hat{\textbf{e}}_z+ \Theta(t)(\mu_e-\mu_g)\hat{\textbf{e}}_z,
\end{align}
where $\Theta(t)$ is the unit Heaviside function. Then, the reaction field in Eq.\ (\ref{eq3.3.2}) becomes 
\begin{align}
\label{eq3.3.4}
\textbf{R}(t)&=\frac{1}{a_0^3}\chi_{s} \mu_g \hat{\textbf{e}}_z + \frac{1}{a_0^3} \int_{-\infty}^t d t' \bigg[\chi(t-t')-\frac{3a_1}{a_0}\exp[-\alpha t]\chi(t-t')\bigg] \Theta(t')\Delta\mu \hat{\textbf{e}}_z,
\end{align}
where $\chi_{s}=\int_{-\infty}^t dt' \chi(t-t') =  \frac{2(\epsilon_s-1)}{2\epsilon_s+1}$ and $\Delta\mu=\mu_e-\mu_g$.
The first term in Eq.\ (\ref{eq3.3.4}) describes the static reaction field before excitation when the solvent is in equilibrium with the ground state dipole moment $\boldsymbol{\mu}_g$. The second term is the change of the reaction field  after the sudden change of the dipole moment to $\boldsymbol{\mu}_e$ which the solvent has to readjust to. In addition, due to the excitation, the molecular radius begins to shrink from its initial value $a_0+a_1$ to $a_0$, which gives rise to an additional explicit time-dependent contribution given by the third term. 

At time $t$, the solute suddenly fluoresces and reaches again its ground state such that also its dipole moment goes back to $\boldsymbol{\mu}_g$. The solvent immediately reacts in the continuum's approach with the fast (or optical) contribution $\epsilon_\infty$. As we set this value to $\epsilon_\infty =1$, there is no further contribution to the reaction field coming from the sudden dipole change arising from $\chi_\infty=2a(t)^{-3}(\epsilon_\infty-1)(2\epsilon_\infty+1)^{-1}$.

The difference of the solvation energies  between the excited state and the ground state of the molecule at time $t$ thus readily follows as 
\begin{align}
\label{eq3.3.5}
\Delta E_{solv}(t)=&-\boldsymbol{\mu}_e\textbf{R}(t)+\boldsymbol{\mu}_g\textbf{R}(t) \\ \notag
=& -\frac{1}{a_0^3}\mu_g \chi_s \Delta \mu - \frac{\Delta\mu^2}{a_0^3}  \int_{-\infty}^t d t' \bigg[\chi(t-t')-\frac{3a_1}{a_0}\exp[-\alpha t]\chi(t-t')\bigg] \Theta(t') \\ \notag
=& -\frac{1}{a_0^3}\mu_g \chi_s \Delta \mu + \Delta E (t),
\end{align}
where only the second term is time-dependent such that the first term cancels out in determining $S(t)$ of Eq.\ (\ref{eq3.3.1}). 
Next, we use the complex Fourier transform $f(z)=\int_{-\infty}^\infty dt \, e^{-izt} f(t)=\mathcal{F}[f(t)]$, where $f(z)$ is analytic for $\rm{Im}(z)<0$ and $f(t\to \infty) < \infty$. Applied to $\Delta E(t)$ and by using the convolution theorem and the Fourier transform of the Heaviside function $\mathcal{F}[\Theta(t)]=1/iz$, we find
\begin{align}
\label{eq3.3.6}
\Delta E(z)=-\frac{\Delta\mu^2}{a_0^3} \bigg[\frac{\chi(z)}{iz}-\frac{3a_1}{a_0}\frac{\chi(z-i\alpha)}{i(z-i\alpha)}\bigg],
\end{align}
where $z=\omega-i\eta$ with $\omega$ real and $\eta$ representing a small positive number \cite{hsu1997}. 
The inverse transform of Eq.\ (\ref{eq3.3.6}) for $t\geq0$ leads to
\begin{align}
\label{eq3.3.7}
\Delta E(t)=&-\frac{\Delta\mu^2}{2\pi a_0^3} \int_C dz e^{izt}\bigg[\frac{\chi(z)}{iz}-\frac{3a_1}{a_0}\frac{\chi(z-i\alpha)}{i(z-i\alpha)}\bigg] \\  \label{eq3.3.8}
=&-\frac{4\Delta\mu^2}{\pi a_0^3} \int_0^{\infty} d\omega \frac{\cos[\omega t]}{\omega}\Im\bigg[\frac{\epsilon(\omega)-1}{2\epsilon(\omega)+1}\bigg]-\frac{\Delta\mu^2 \chi_s}{a_0^3}\bigg[1-3\frac{a_1}{a_0}e^{-\alpha t}\bigg] \\ 
\label{eq3.3.9}
=&\frac{\Delta \mu^2}{a_0^3}\chi_s\bigg\{ e^{-\omega_D t}-\bigg[ 1-3\frac{a_1}{a_0}e^{-\alpha t} \bigg]\bigg \}. 
\end{align}
The contour $C$ of integration in Eq.\ (\ref{eq3.3.7}) is a path parallel to but slightly below the real axis in the complex plane. Further details for calculating the first term in Eq.\ (\ref{eq3.3.7}) are given in the Supporting Information and in Ref.\ \cite{hsu1997} . Moreover, we apply the theorem of residues to the second term for the singularity at $z=i\alpha$. We use the Debye form of the dielectric function  $\epsilon(\omega)=1+\frac{\epsilon_s-1}{1+i\omega \tau_D}$ to evaluate the integral in Eq.\ (\ref{eq3.3.8}). The resulting form $\omega_D=(2\epsilon_s+1)/(3\tau_D)=\tau_L^{-1}$ in Eq.\ (\ref{eq3.3.9}) is the inverse dipolar longitudinal relaxation time \cite{bag1984}.

By this, we obtain the final result of the time-dependent fluorescence Stokes shift according to Eq.\ (\ref{eq3.3.1}) in the form 
\begin{align}
\label{eqStokes}
S(t)&=\frac{1}{1+Q} e^{-\omega_D t} + \frac{Q}{1+Q} e^{-\alpha t} \\ 
& \simeq (1-Q )e^{-\omega_D t} + Q e^{-\alpha t}  \, , \notag
\end{align}
where $Q=3\frac{a_1}{a_0}$.

\section{Results}
\label{StokesResults}
\begin{figure}[h!!!] 
 \begin{center}    
   \includegraphics[width=\textwidth]{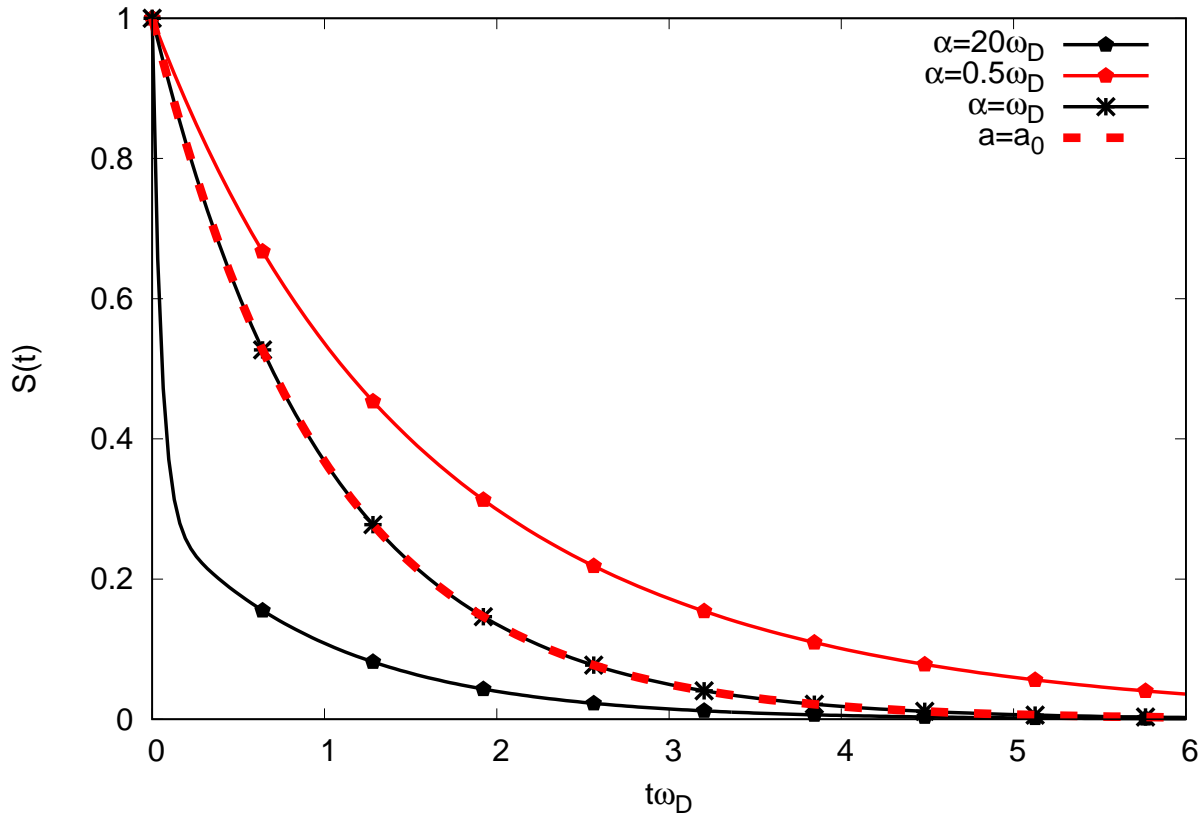}
  \caption{\label{figStokes} Nonequilibrium time-dependent fluorescence Stokes shift due to a suddenly changing dipole moment in an shrinking Onsager cavity for different shrinking rates $\alpha$ for  $a_1=0.01a_0$. The dashed red line shows the time-dependent Stokes shift in a static Onsager cavity of constant radius $a_0$.}
\end{center}
\end{figure}
The nonequilibrium time-dependent Stokes shift of Eq.\ (\ref{eqStokes}) reveals a bi-exponential decay where the first term reflects the collective bulk water relaxation while the second term describes the motional changes at the solute-solvent interface associated with a possible self-motion of the solute and the corresponding shrinking rate $\alpha$.  Interestingly, also the first term related to the bulk water relaxation carries information about the time-dependence of the solute via the prefactor proportional to $Q$.  
Fig.\ \ref{figStokes} shows the nonequilibrium time-dependent Stokes shift for different shrinking rates, which are equal to (black-star line), larger (black-diamond line) or smaller (red-diamond line) than the inverse longitudinal relaxation time. For $\alpha > \omega_D$, the bulk contribution dominates and the rapidly decaying contribution of the solute motion is almost negligible. When 
$\alpha < \omega_D$, the long-term tail of $S(t)$ is dominated by the solute's motion and the bulk contribution has decayed rapidly. For the special case $\alpha = \omega_D$, the two contributions cannot be separated in the Stokes shift.

\begin{figure}[h!!!] 
 \begin{center}    
   \includegraphics[width=\textwidth]{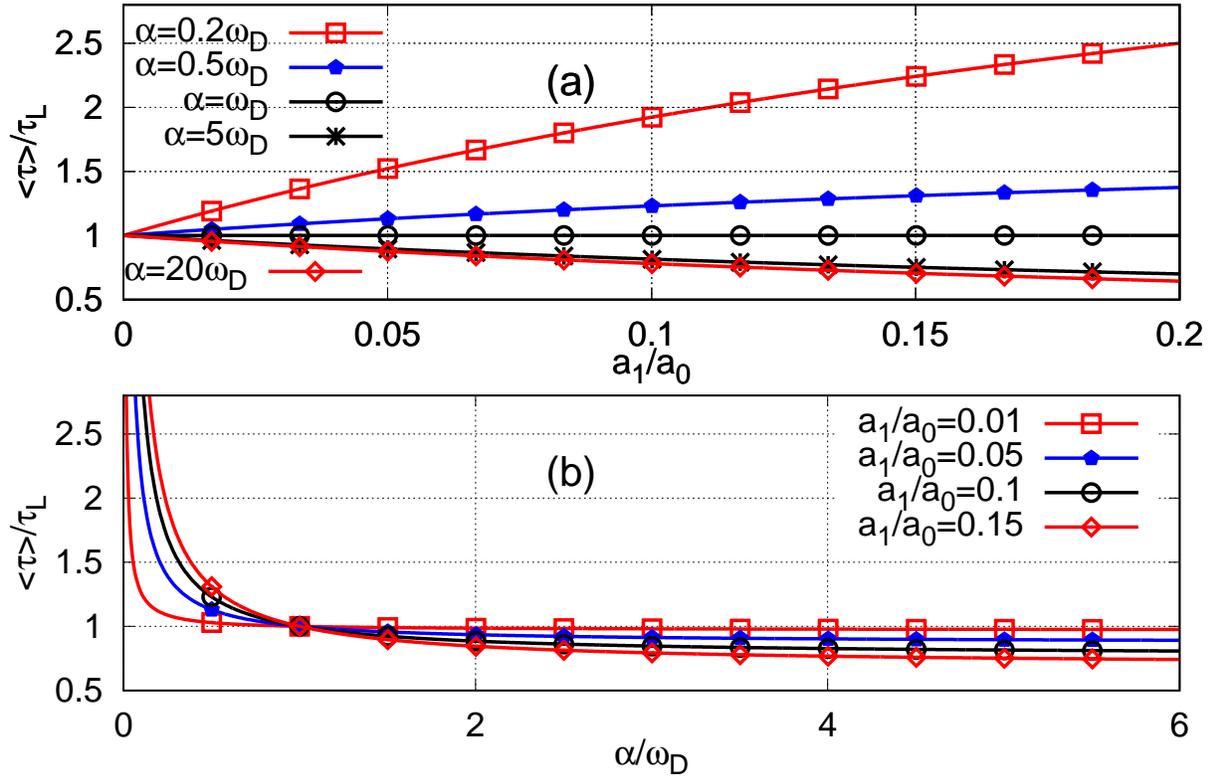}
  \caption{\label{figStokes2} (a) Average relaxation time $\langle \tau \rangle$ for the time-dependent fluorescence Stokes shift  in dependence of the relative radial change $a_1/a_0$ for different shrinking rates $\alpha$. (b) Average relaxation time $\langle \tau \rangle$ in dependence of the shrinking rate $\alpha$ for different relative radial changes $a_1/a_0$.}
\end{center}
\end{figure}

Next, we determine the average relaxation time 
\begin{align}
\label{eqavtime}
\langle \tau \rangle & =\frac{1}{1+Q}\tau_L+\frac{Q}{1+Q} \alpha^{-1}  \\ 
& \simeq (1-Q )\tau_L + Q  \alpha^{-1}  \, , \notag
\end{align}
for the time-dependent Stokes shift of Eq.\ (\ref{eqStokes}), where the prefactors sum up to $1$. Fig.\ \ref{figStokes2} (a) shows  different modes of behavior for the average relaxation time with relative changes of the Onsager radius $a_1/a_0$. For $\alpha<\omega_D$, the shrinking is slower than the longitudinal relaxation  which highly increases the average relaxation time in comparison to bulk water relaxation determined by $\tau_L$. In this regime, the shrinking exceeds the conventional water relaxation process and becomes directly measurable via the slow long-time decay of the time-dependent fluorescence spectrum. With a larger absolute change of the Onsager radius $a_0+a_1\to a_0$, the average relaxation time grows further because the shrinking process is noticeable for longer times. 
When $\alpha=\omega_D$ the average relaxation time shows the bulk water property $\tau_L$ although a shrinking occurs. In this special case, the interfacial solute-solvent motion remains undisclosed. 
A faster shrinking for $\alpha>\omega_D$ leads to a smaller average relaxation time in comparison to bulk water. Now, the fast Onsager radial motion reduces the overall relaxation which decays further with relative radial change $a_1/a_0$. 
Interestingly enough, the relative impact of a slowly shrinking Onsager cavity on the average relaxation time is more pronounced  than for a rapidly shrinking sphere. Fig.\ \ref{figStokes2} (b) shows the diverging increase of $\langle \tau \rangle$ for small shrinking rates $\alpha\ll \omega_D$. Such a behavior can in principle be recorded by a long-tail decay in the associated fluorescence spectrum which ranges up to several hundred ps \cite{pal2004}. Thus, the slow solute-solvent interfacial motion by, e.g., self-motion of a protein can become  directly detectable. For the fast shrinking $\alpha >\omega_D$, the average relaxation time approaches quickly its steady-state value $\langle \tau \rangle\simeq (1-Q) \tau_L$. In this regime, the faster shrinking has no further impact on the relaxation time and one can consider the fluctuating solvent to be "quasi-instantaneously" close to the solute. It is evident, that this slightly reduces the average relaxation time, see also Refs.\ \cite{nal2014,kir2018}.  

\section{Discussion}

An example for a bimodal experimental fit to the measured time-dependent fluorescent Stokes shift is given by the dye Hoechst 33258 in solution with DNA . The data show clear evidence of different motional contributions to relaxation. The experimental data for the Stokes shift confirm that relaxation due to the fluctuations in the bulk water solvating the dye bound to DNA is slowed down,  but contributes to the fast relaxation times ($0.2$ ps and $1.2$ ps). On the other hand,  the DNA self-motion, which occurs on a time scale of $\sim$20 ps, modifies the long-time components ($1.4$ ps and $19$ ps) of the solvation response \cite{fur2008,fur2010}. 

We have shown that the standard Onsager continuum model for the time-dependent Stokes shift can be generalized to a generic model which includes an explicit time-dependent motion in the interfacial solute-solvent region. By this, we are able to generalize the Bagchi-Oxtoby-Fleming theory towards nonequilibrium. The change in the solute radius upon photoexcitation drives nearby solvent molecules  into locations that are never occupied at equilibrium. Clearly,  the proposed dielectric model is still phenomenological, but provides an intuitive picture of the involved contributions to the relaxation dynamics. Our theory can, in principle, be extended to cases where the solute has different geometrical shapes such as spheroids or where the solvent is a mixture of polar solvent molecules by adapting dielectric constants. Furthermore, the dielectric solvent incorporates important solute-solvent interactions which are commonly neglected in molecular modeling \cite{bag2010}. The explicit inclusion of electronic and collective polarization in MD simulations, for example, is numerically expensive, especially from the point of view of the actual accurate force field parametrization \cite{kirb2019}. The used continuum model incorporates long-range electrostatic interactions as well as polarization and, in addition, gives a configurationally sampled solvent effect avoiding delicate statistical sampling averages. The frequency-dependent dielectric function entering in fluorescence Stokes shift in Eq.\ (\ref{eqStokes}) automatically accounts for time averages \cite{men2010}.
In addition, microscopic models where the motion of individual molecules is incorporated may be more accurate, but only for the specific situation under investigation. Yet, molecular dynamics simulations have successfully examined the origin of the Stokes shift  and have revealed numerically that the bulk water component dominates at short times but rapidly decays. The long-time behavior is dominated by the solute dynamics, i.e., mostly the self-motion of a dissolved protein \cite{nil2005}, such as DNA solvation dynamics \cite{muk2018}. 
The proposed nonequilibrium continuum model clearly reveals this mechanism in a simple and transparent manner and agrees with this observation. It also confirms experimental measurements of fluorescence Stokes shifts which show two (or more) components of  decaying exponentials with different time scales involved. 

On the whole, different spectroscopic decay times may be assigned to specific motions in a more complex system. However, the complexity of the relaxation channels and their interplay in a macromolecule in solution make it often hard to interpret the various contributions to the time-dependent Stokes shift. In the study of the hydration dynamics, for example, one can attribute the dynamic exchange of 'bound' and 'free' water molecules between the hydration shell and the bulk water to a slow relaxational component, but also to self motion of the protein or its side chains. Also, a coupled motion of both dynamical effects could occur \cite{bag2010}. Our proposed time-dependent continuum model explicitly confirms the coupling of different motional contributions related by the prefactor $Q$ where the time-dependent configurational changes of the solute enters. In total, we suggest a combination of exact microscopic modeling of distinct solute-solvent motions and explicit dynamical dielectric continuum models to team the strengths of both. Then, solvation dynamics and the resulting solvent effects in physical and chemical processes in liquids can be further elucidated even in highly complex biological systems like proteins. 

\section*{Acknowledgments}
This work is supported by the DFG-Sonderforschungsbereich 925 "Light-induced dynamics and control of correlated
quantum systems" (A4, project  number 170620586). 

\section*{Supporting Information}
The Supporting Information includes the detailed calculation of the first term in Eq.\ (\ref{eq3.3.7}).


%
%

%



\end{document}